\documentclass[epj]{svjour}
\usepackage{graphicx}
\newcommand{\eps}{\epsilon}
\newcommand{\om}{\omega}
\newcommand{\si}{\sigma}
\newcommand{\ra}{\rightarrow}
\newcommand{\beq}{\begin{equation}}
\newcommand{\eeq}{\end{equation}}
\newcommand{\beqa}{\begin{eqnarray}}
\newcommand{\eeqa}{\end{eqnarray}}
\newcommand{\bea}{\begin{array}}
\newcommand{\eea}{\end{array}}
\newcommand{\bite}{\begin{itemize}}
\newcommand{\eite}{\end{itemize}}
\newcommand{\bfig}{\begin{figure}}
\newcommand{\efig}{\end{figure}}
\newcommand{\btable}{\begin{table}}
\newcommand{\btabul}{\begin{tabular}}
\newcommand{\etabul}{\end{tabular}}
\newcommand{\etable}{\end{table}}
\newcommand{\ben}{\begin{enumerate}}
\newcommand{\een}{\end{enumerate}}

\begin{document}

\title{Atomic density of a harmonically trapped ideal gas near Bose-Einstein transition temperature}
\author{R. Hoppeler\inst{1} \and J. Viana Gomes\inst{1} \thanks{\emph{Permanent address: Departamento
de Fisica, Universidade do Minho, Campus de Gualtar, 4710-057
Braga, Portugal}} \and  D. Boiron\inst{1}
}                     
%
%
\institute{Laboratoire Charles Fabry de l'Institut d'Optique,\\
UMR 8501 du CNRS, F-91403 Orsay Cedex, France}
\date{Received: date / Revised version: date}
\mail{denis.boiron@iota.u-psud.fr}

\abstract{We have studied the atomic density of a cloud confined
in an isotropic harmonic trap at the vicinity of the Bose-Einstein
transition temperature. We show that, for a non-interacting gas
and near this temperature, the ground-state density has the same
order of magnitude as the excited states density at the centre of
the trap. This holds in a range of temperatures where the
ground-state population is negligible compared to the total atom
number. We compare the exact calculations, available in a harmonic
trap, to semi-classical approximations. We show that these latter
should include the ground-state contribution to be accurate.
} 
%
\maketitle

The phenomenon of Bose-Einstein condensation (BEC) is a phase
transition. Below the critical temperature $T_c$, the ground-state
population, which is the order parameter, becomes macroscopic.
This phenomenon, that happens strictly speaking only at the
thermodynamic limit, is usually illustrated in textbooks with a
homogeneous gas. Experimentally, the Bose-Einstein condensation of
dilute gases has been observed since 1995 with atoms confined in a
harmonic trap \cite{Varenna}. These stimulating experimental data
have quickly pointed out that two effects had to be taken into
account: the interatomic interactions and the finite number of
atoms \cite{stringari-revue}. Several papers, as the present one,
have studied harmonically trapped ideal gases containing a finite
number of atoms. Two quantities have been investigated in detail:
the atom number
\cite{grossmann,ketterle,stringari,darnval,giorgini,pathria} and
the specific heat \cite{inc,darnval,pathria}. For a finite but
large (typically $10^6$) number of atoms, the properties of the
atomic cloud change abruptly at a characteristic temperature we
will name the transition temperature $T^*$. This temperature is
shifted compared to $T_c$, but by a small amount, typically of few
percent for atom numbers around $10^6$. There is also a
characteristic temperature for the specific heat; it is different
from the previous one but still close to $T_c$ \cite{inc,pathria}.

Surprisingly, less attention has been paid on the atomic density
of an ideal gas \cite{krauth96}. In a homogeneous gas it is
obviously equivalent to the atom number but this is no more the
case in a spatially varying potential. It becomes the good
parameter of the theory, in particular to perform local density
approximations. This quantity is then particularly important for
the study of the shift of the critical temperature by the
interatomic interactions, both within the mean-field approximation
\cite{stringari} and beyond this approximation \cite{arnold2}. We
will show, in the case of an isotropic harmonic trapping and for a
finite atom number, that the ground-state density at the centre of
the trap increases much more sharply than its population as the
temperature decreases. This leads to the fact that near the
Bose-Einstein transition temperature the density is already
dominated by the ground-state contribution. This holds whatever
the atom number is, and is a remanence of the infinite
compressibility of an ideal gas at the thermodynamic limit
\cite{compressibilite}. Usual semi-classical approximations do not
take into account the ground-state contribution and then fail in
the vicinity of the Bose-Einstein transition temperature. This is
not a finite size effect in the sense that it is not related to
the discretization of the excited states energy levels. We will
compare the exact results with semi-classical approximations. The
addition of the ground-state contribution on the latter ones
improves their accuracy. We will finally show that the influence
of the ground-state is smaller if the measured quantity is the
density integrated over at least one dimension. It
is still large for typical experimental parameters.\\

We will perform our calculations in the grand canonical ensemble
(GCE). Then, the Bose-Einstein distribution gives the population
$N_i$ of a given energy level $\eps_i$: $
N_i=(e^{\beta(\eps_i-\mu)}-1)^{-1}$ with $\sum\limits_{i=0}^\infty
N_i=N$. Here $\beta=1/k_BT$ with $k_B$ the Boltzmann's constant,
$\mu$ the chemical potential and $N$ the total atom number. The
equivalence between GCE and the canonical or microcanonical
ensemble, these latter being probably more appropriate
descriptions, is generally not guaranteed, especially for systems
that are not at the thermodynamic limit. For instance, it is well
known that the GCE predicts unphysical large fluctuations of the
condensate population at low temperature \cite{landau}. However,
the authors of Ref. \cite{krauth96,Politzer96,Olshanii97} have
shown that the occupation numbers $N_i$ in GCE are very close to
the ones in the canonical ensemble. The difference is more
pronounced for small atom number and anisotropic clouds. As a
result and because GCE enables to give analytic expressions on
contrary to the other ensembles, we will use GCE in the following.

For a fixed atom number, the chemical potential increases as the
temperature decreases. As $\mu$ has to be smaller than $\eps_0$,
the ground-state energy, the excited states population will
saturate when $\mu$ approaches $\eps_0$ whereas $N_0$ is still
increasing: $N-N_0=\sum\limits_{i=1}^\infty N_i(\mu,T)\le
\sum\limits_{i=1}^\infty N_i(\eps_0,T)$. As in Ref.
\cite{stringari-revue,castin}, we will define the transition
temperature $T^*$ as the temperature for which the excited states
saturated population is equal to the total atom number:
\begin{equation}\sum\limits_{i=1}^\infty N_i(\eps_0,T^*)=N\label{def_T*}\end{equation}

As pointed out in the introduction, there is not a unique
definition of the transition temperature for a finite atom number.
Other definitions use, for instance, a change in the slope for the
condensate fraction in function of temperature (more explicitly
$\frac{d^3(N_0/N)}{dT^3}=0$) \cite{Bergeman}, a change in the
power dependence on the condensate fraction in function of the
atom number \cite{pathria}, which are also pertinent. We have
checked that these various definitions affect marginally the value
of $T^*$ and do not modify our conclusions \cite{resTc}. In the
following we will then use Eq.(\ref{def_T*}) to define $T^*$. Note
that the chemical potential $\mu^*$ at the transition temperature
is close but not equal to the ground-state energy; it is
determined by the constraint
\begin{equation}\sum\limits_{i=0}^\infty
N_i(\mu^*,T^*)=N\label{def_mu*}\end{equation}

There are only a few examples of trapping potentials where the
eigen-energies and the eigen-functions are known exactly.
Semi-classical approximations give usually accurate enough results
and are suited to include interatomic interactions, at least
perturbatively. We will derive various type of semi-classical
approximations in the following and test their accuracy because
the harmonic potential is an exactly solvable potential.

\begin{figure}
\begin{center}
\includegraphics[scale=0.6]{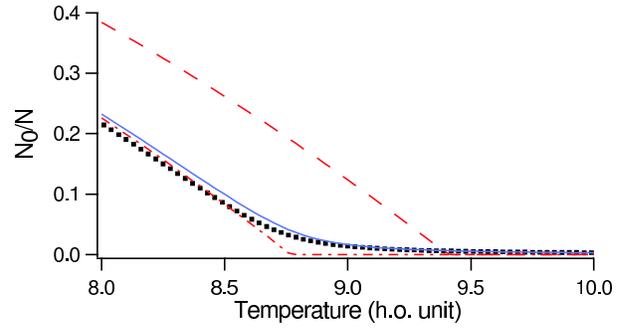}
\end{center}
\caption{Ground-state population fraction in function of the
temperature in $\hbar\om/k_B$ unit for a cloud of $10^3$ atoms.
The dotted curve corresponds to the exact result given by model
$ex$. The solid, dot-dashed and dashed lines correspond
respectively to the semi-classical models $sc$, $sc_0$ and
$sc_\infty$. The last two neglect the ground-state contribution
above their corresponding transition temperature and the first two
take into account finite size effects. The model $sc$ is the
closest to $ex$ near Bose-Einstein transition.}
 \label{frac0}
\end{figure}

We will first examine the situation where $\hbar\om\ll k_BT$ with
$\om$ the oscillation frequency of the isotropic harmonic trap.
This corresponds to the large atom number limit and semi-classical
approximations should work. 
Replacing the discrete energy spectrum by a continuous one and
neglecting the ground-state energy $\eps_0$, the density is
$\rho(r)={1\over\lambda^3}g_{3\over 2}[z\exp(-{\tau\over 2}
(r/\si)^2]$ with $z=e^{\beta\mu}$ the fugacity,
$\tau={\hbar\om\over k_BT}$ and $g_{3\over 2}()$ a Bose function
\cite{bose}. With the above notation, the thermal de Broglie
wavelength is $\lambda=\si\sqrt{2\pi\tau}$ and the size of the
cloud is $\sqrt{{k_BT\over m\om^2}}=\si/\sqrt{\tau}$. Similarly,
the atom number is $N=g_{3}(z)/\tau^3$. Equation~(\ref{def_T*})
leads then to $N=\zeta(3)/{\tau^*}^3$, with $\tau^*$ the value of
$\tau$ at $T=T^*$. The above expressions for the density and atom
number are in fact approximations for the excited states and do
not contain the ground-state contribution. Then $\mu^*$ defined by
Eq.(\ref{def_mu*}) is equal to 0 and $z^*=1$. The transition
temperature defined here corresponds to the critical temperature
$T_c$. The peak density at the transition temperature is then
given by $\rho(0)\lambda^3=g_{3\over 2}(z^*)=\zeta(3/2)\approx
2.612$. For temperatures below $T_c$, the excited states
population is given by $\zeta(3)/\tau^3$. Then, the ground-state
population fraction is $N_0/N=0$ for $T>T_c$ and
$N_0/N=1-(T/T_c)^3$ for $T<T_c$. This fraction will be plotted in
fig.\ref{frac0}, labelled with $sc_\infty$.

These approximations are too crude and give inaccurate results for
the atomic density, however. The reason is that the ground-state
contribution cannot be neglected. A better expression is
$\rho(r)={1\over\lambda^3}g_{3\over 2}[z e^{-{\tau\over 2}
(r/\si)^2}] + \rho_0(r)$ and similarly $N={1\over\tau^3}g_{3}(z)+
N_0$ with $\rho_0(r)={N_0\over (\sqrt{\pi} \si)^3}e^{-(r/\si)^2}$
and $N_0={z\over 1-z}$. The value of $T^*$ is unchanged as it is
defined by the excited states saturation, but $z^*$ is now
different from 1. Using $g_3(z^*)\approx \zeta(3)-\zeta(2)x^*$
with $z^*=e^{-x^*}$ ($x=\beta(\eps_0-\mu)>0$), one finds using
Eq.(\ref{def_mu*}) that $x^*\approx{\tau^*}^{3/2}/\sqrt{\zeta(2)}$
\cite{pathria}. The ground-state population is $\sim 1/x^*$ and,
as expected, is vanishingly small as $\tau^*\rightarrow 0$
compared to the excited-state population $\zeta(3)/{\tau^*}^3$.
The ground-state peak density is $\sim {1\over (\sqrt{\pi} \si)^3
x^*}$ whereas the excited state peak density is
$\zeta(3/2)/{\lambda^*}^3$. As $\lambda^* = \si\sqrt{2\pi\tau^*}$,
the two quantities have the same order of magnitude! The above
high-N analysis predicts then that the degeneracy parameter at the
transition temperature is
$\rho(0)\lambda^3=\zeta(3/2)+2\sqrt{2\zeta(2)}\approx 6.24$ and
not 2.612. The ground-state population is extremely small but the
size of its wave-function is also extremely small compared to the
atomic cloud size. For a harmonic trap both depend on the same
small parameter, raised to the same power. So, even for very large
atom number, the traditional criterion for BEC should be modified.
This effect is linked to the pathological behaviour of the
ground-state density at the thermodynamic limit, i.e. the infinite
compressibility of an ideal gas \cite{compressibilite}. This limit
means $N\ra\infty$ with $N\om^3\ra$ constant. The ground-state
size being $\si=\sqrt{\hbar/m\om}$, the density of that state
behaves as $\sqrt{N}$ below threshold and is then infinite at the
thermodynamic limit whereas the density above $T_c$ is finite.

We will now address the case of atom numbers in the accessible
experimental range, $10^3-10^6$. It is well known that the
transition temperature will be shifted compared to $T_c$
\cite{grossmann,ketterle,darnval}. A better approximation, which
takes into account the ground-state energy to first order, is
$\rho(r)={1\over\lambda^3}\{g_{3\over 2}[\tilde z(r)]+{3\tau\over
2}g_{1\over 2}[\tilde z(r)]\}$ where $\tilde z(r)= z
e^{-{\tau\over 2} (r/\si)^2}$. Then
$N={1\over\tau^3}[g_{3}(z)+{3\tau\over 2}g_{2}(z)]$.  The
corresponding transition temperature is $T^*_{sc}$ such that
$N={1\over\tau^{* 3}_{sc}}[\zeta(3)+{3\over
2}\zeta(2)\tau^*_{sc}]$. This is the usual semi-classical
approximation found in the literature. The ground-state population
fraction is then $N_0/N=0$ for $T>T^*_{sc}$ and $N_0/N=1-({T\over
T^*_{sc}})^3\frac{\zeta(3)+{3\tau\over
2}\zeta(2)}{\zeta(3)+{3\tau^*_{sc}\over 2}\zeta(2)}$ for
$T<T^*_{sc}$. This fraction, also plotted in fig.\ref{frac0}, will
be labelled with $sc_0$. Note that $g_{1\over 2}(z)$ diverges at
$z=1$ \cite{Yukalov05}, meaning that this approximation is
intrinsically inaccurate near the centre of the trap and near the
transition temperature. This divergence is however weak, and any
spatial integration would give a finite result. We can still cure
this pathology by adding, as before, the ground-state
contribution. We obtain then
\begin{equation}
\left\{\!\!\!\btabul{l} $\rho_{sc}(r)={1\over\lambda^3}\{g_{3\over
2}[\tilde z(r)]+{3\tau\over 2}g_{1\over 2}[\tilde z(r)]\}
+ {z\over 1-z}{e^{-({r\over\si})^2}\over (\sqrt{\pi} \si)^3}$\\
$N={1\over\tau^3}[g_{3}(z)+{3\over 2}\tau g_{2}(z)]+{z\over
1-z}$\label{equations semi-classique}\\
$T^*_{sc}$ such that $N={1\over\tau^{* 3}_{sc}}[\zeta(3)+{3\over
2}\zeta(2)\tau^*_{sc}]$ \etabul\right.
\end{equation}

This semi-classical approximation will be labelled with $sc$ in
the following. The comparison of $T^*_{sc}$ with the value given
by the exact model (see below) can be used to check the finite
size correction. Even so, this comparison is useless to check the
contribution coming from the ground state since it does not depend
on it (same transition temperature as $sc_0$).

We can now test these semi-classical approximations for a
harmonically trapped gas. As we referred before, for this case,
the eigen-energies and the eigen-functions are known exactly. The
corresponding expressions of the atomic
density and atom number \cite{landau}, labelled with $ex$ in the following, are :\\

$\left\{\btabul{l}
$\rho_{ex}(r)={1\over(\sqrt{\pi}\si)^3}\sum\limits_{l=1}^\infty
{z^l\over(1-e^{-2\tau l})^{3/2}}
\;e^{-\tanh({\tau l\over 2})({r\over\si})^2}$\\
$N=\sum\limits_{l=1}^\infty {z^l\over(1-e^{-\tau l})^3}$\\
$T^*_{ex}$ such that $N=\sum\limits\limits_{l=1}^\infty ({1\over(1-e^{-\tau^*_{ex} l})^3}-1)$ \etabul\right.$\\
where, here $z=e^{\beta(\mu-\eps_0)}$. The semi-classical model
corresponds to a Taylor expansion in
$\tau$ of these last expressions.\\

\begin{figure}
\begin{center}
\includegraphics[scale=0.6]{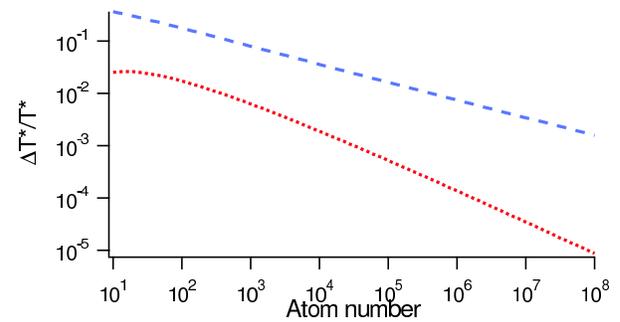}
\end{center}
\caption{Relative shift of the semi-classical transition
temperatures $T_c$ (dashed line) and $T_{sc}^*$ (dotted line) to
$T_{ex}^*$ (see text) in function of the atom number. Both
temperatures converge for high atom numbers. The critical
temperature at thermodynamic limit, $T_c$, deviates by less than
$1\%$ for $N>5\,10^5$. The semi-classical transition temperature
defined for a finite atom number, $T_{sc}^*$, is much more
accurate and deviates by less than $1\%$ for $N>400$.} \label{tc}
\end{figure}

In fig.\ref{frac0} we plot the ground-state population fraction in
function of the temperature for the various models described
above. When the number of atoms is only $10^3$, finite size
effects are large. The prediction of model $sc_\infty$ is clearly
wrong compared to the exact model prediction. On contrary models
$sc_0$ and $sc$ give a result close to the one of $ex$
\cite{calcul}. Figure \ref{tc} shows the relative deviations of
$T_c$ and $T^*_{sc }$ from $T^*_{ex}$ in function of the atom
number. As expected the different values are similar but, as
above, the model $sc$ give a closer result to $ex$ than
$sc_\infty$. The value $T^*_{sc}$ deviates less than $1\%$ for
$N>400$ and the relative shift is $\sim 10^{-4}$ for typical
experimental atom numbers. This is well below actual experimental
uncertainties. The thermodynamic value $T_c$ deviates more,
typically 1~\% but is still close to $T_{ex}^*$
\cite{grossmann,ketterle,darnval,pathria}. The discrepancy with
$T_c$ would have been more pronounced for an anisotropic trap (see
below).

This two figures illustrate what is called finite size effects,
the fact that the energy level spacing is not negligible compared
to the temperature. What we are interested in is the role of the
ground-state. For this, the transition temperature and the
condensate population fraction are not the best observables. It is
nevertheless already clear from fig.\ref{frac0} that $sc$ is a
significant improved model to describe semi-classically a cloud
near degeneracy compared to $sc_0$. The high-N model predicts that
the ground-state influence should be much more pronounced on the
peak density. We will now focus our attention on that observable,
only in the more pertinent comparison between the models $sc$ and
$ex$.

\begin{figure}
\begin{center}
\includegraphics[scale=0.6]{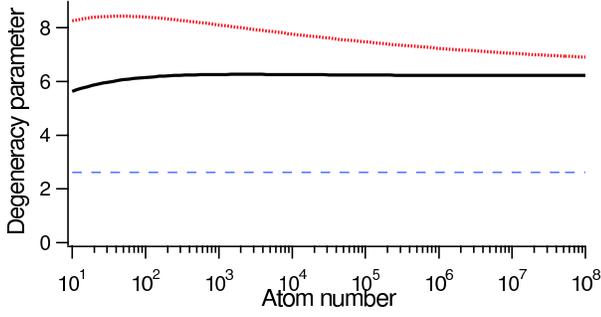}
\end{center}
\caption{Degeneracy parameter $\rho(0)\lambda^3$ in function of
the atom number $N$ for clouds at the transition temperature. The
dotted line corresponds to the semi-classical model $sc$ at
$T=T_{sc}^*$ and the solid line to model $ex$ at $T=T_{ex}^*$.
Even if the degeneracy parameters are somewhat different, they
both differ significantly to the usual value of 2.612 (dashed
horizontal line). This deviation is due to an under-estimation of
the ground-state density. The actual values are close to our
high-N prediction of 6.24 (see text). } \label{phase}
\end{figure}

This is first illustrated on fig. \ref{phase} where the degeneracy
parameter $\rho(0)\lambda^3$ is plotted in function of the atom
number for clouds at $T=T^*$. We plot this number for the
semi-classical approximation $sc$ and for the exact model, $ex$.
The two curves are higher than $2.612$. This highlights the
inaccuracy of the standard semi-classical models ($sc_0$ or
$sc_\infty$) that do not take into account the ground-state
contribution. It confirms also the calculation developed above.
The degeneracy parameter is astonishingly constant till $10^3$
atoms and does not differ much even for smaller atom numbers.
Models $sc$ and $ex$, which have almost the same transition
temperature, have the same asymptotic value of the degeneracy
parameter. This value, $6.24$, is the one predicted by our high-N
analysis. The model $sc$ is significantly higher than this value
for experimentally accessible atom numbers. This is because our
first analysis does not take into account the ${3\over 2}\tau$
term of model $sc$. To first order \cite{bose}, $x^*_{sc
}\approx{(\tau_{sc }^*)^{3\over 2}\over\sqrt{\zeta(2)}}(1+{9\over
8\zeta(2)}\tau^*_{sc}\ln\tau^*_{sc })$  and is then slightly
smaller than $(\tau_{sc }^*)^{3\over 2}\over\sqrt{\zeta(2)}$.
Consequently the ground-state peak density is bigger at $T^*_{sc}$
using model $sc$ than at $T_c$ using the high-N model. The excited
states peak density is also higher in model $sc$ because of this
${3\over 2}\tau g_{1\over 2}$ term.

\begin{figure}
\begin{center}
\includegraphics[scale=0.6]{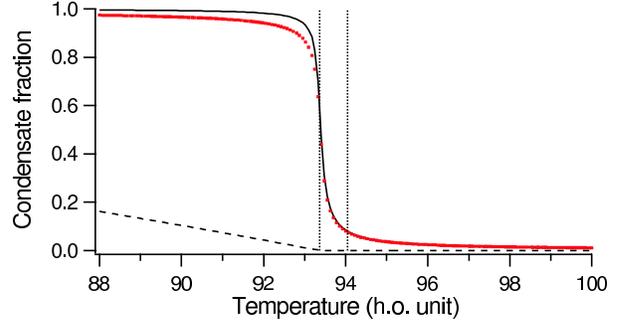}
\end{center}
\caption{Condensate atom number
 fraction $N_0/N$ (dashed line) and peak density fraction $\rho_0(0)/\rho(0)$ (solid line) in
 function of the temperature in harmonic oscillator unit $\hbar\om/k_B$, using model $ex$. The cloud contains $10^6$ atoms.
 The transition temperature is $T_{ex}^*=93.37\hbar\om/k_B$
 and the asymptotic thermodynamic temperature is
 $T_c=94.05\hbar\om/k_B$. The positions of these temperatures are
 shown as vertical lines in the figure.
 The ground-state peak density increases much more sharply than the
 ground-state population around the transition temperature.
 The former has also a significant value above $T^*_{ex}$.
 The model $sc$ is indistinguishable for $N_0/N$, but is slightly different for $\rho_0(0)/\rho(0)$ (dotted line).}
 \label{frac}
\end{figure}

\begin{figure}
\begin{center}
\includegraphics[scale=0.6]{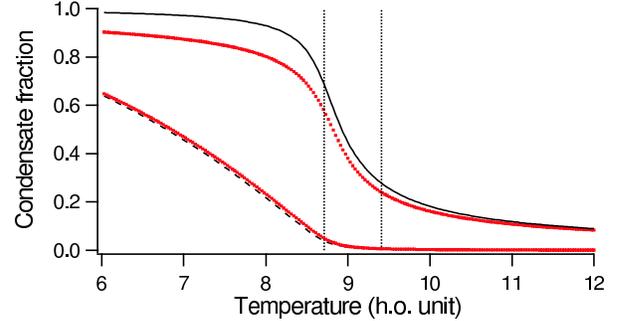}
\end{center}
\caption{Same as in Fig.\ref{frac} but with $10^3$ atoms. The
transition temperature is $T_{ex}^*=8.71\hbar\om/k_B$ and the
asymptotic thermodynamic temperature is $T_c=9.41\hbar\om/k_B$.
Since the number of populated states is considerably reduced
compared to fig.\ref{frac}, the discrepancy between $sc$ (dotted
lines) and $ex$ is more pronounced. This also explains why the
increase of the condensate peak density is slower.}
 \label{frac2}
\end{figure}

\begin{figure}
\begin{center}
\includegraphics[scale=1]{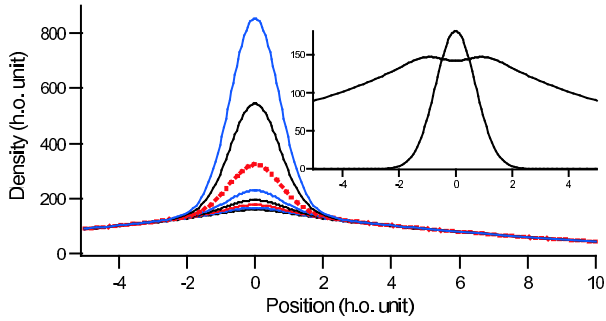}
\end{center}
\caption{Atomic density $\rho_{ex}$ in function of $r/\si$ where
$\si$ is the size of the harmonic oscillator ground-state. The
temperature is $T=93.37\hbar\om/k_B$ and the atom number $N$ spans
from $0.990\;10^6$ to $1.004\;10^6$ by step of $2000$ atoms. The
curve at threshold is in dotted line and corresponds to $10^6$
atoms. The inset shows the excited states and ground state density
profile at threshold. The dip around $r=0$ is mainly due to the
first excited state population.} \label{profilnat}
\end{figure}

The next three figures deal with the cloud properties around the
Bose-Einstein threshold. Figure \ref{frac} and fig.\ref{frac2}
show the evolution of the condensate fraction $N_0/N$ and the
condensate peak density fraction in function of $T$ for two
different atom numbers, $10^6$ and $10^3$. Figure \ref{profilnat}
shows the density profile of clouds near degeneracy. What prevails
in fig.\ref{frac} is the sharp increase of the condensate peak
density compared to the condensate population. Moreover the models
$sc$ and $ex$ give very close results validating our analysis on
the ground-state contribution near degeneracy. This means that the
peak density is a much better marker of the Bose-Einstein
threshold than the atom number. This feature is in fact used
experimentally: the appearance of a small peak over a broad
distribution is the usual criterion to distinguish clouds above or
below the transition temperature. This sharpness also explains why
the value of the peak density is very sensitive to the value of
the temperature (cf. fig.\ref{phase}). Figure \ref{frac} shows
also that, above threshold, the ground-state peak density fraction
decays slowly. This is even more pronounced in fig.\ref{frac2}
where $N=10^3$ instead of $10^6$. It comes from the fact that the
number of populated states is not macroscopic anymore
($k_BT<10\hbar\om$) and then the transition is smoother for
smaller atom number. Once again, the density is a better marker of
degeneracy than the atom number. This figure shows also that the
${3\over 2}\tau$ term and the ground-state contribution make the
model $sc$ still very close to model $ex$, respecting the density
and population fractions, even for $10^3$ atoms.

The above analysis is focused on the peak density i. e. at the
centre of the cloud. Figure \ref{profilnat} shows the total
density profile of clouds, all at the same temperature, but
containing different numbers of atoms around $N^*_{ex}$, the
number of atoms for which $T=T^*_{ex}$ ($N=N^*_{ex}$ corresponds
to the dotted line). This figure simulates somehow an experimental
observation of BEC threshold. Only the central part is sensitive
to the atom number; this corresponds to the condensate growing as
the number of atoms is increased and to the fact that the excited
states are already saturated for these atom numbers. Moreover, by
looking at the graph, one would rather think that the
Bose-Einstein transition occurs for a smaller atom number. This
points out that the definition on the transition temperature based
on an atom number criterion does not fully correspond to the one
based on the atomic density which would be more connected to
experiments. The inset shows the excited states and ground state
density profiles at threshold. The excited states density exhibits
a dip in the centre of the cloud, obviously not present in
semi-classical models (monotonic functions). We check that the
height of the dip is proportional to $1/\tau$ and can almost be
totally attributed to the first excited state population. The aim
of this paper is to show the importance of the ground-state in the
study of non-interacting clouds close to threshold. The inset
reveals that the first excited state density is also largely
under-estimated; it represents $\sim 10~\%$ of the peak density
whereas it contributes only to $\sim 0.1~\%$ of the population.

\begin{figure}
\begin{center}
\includegraphics[scale=0.6]{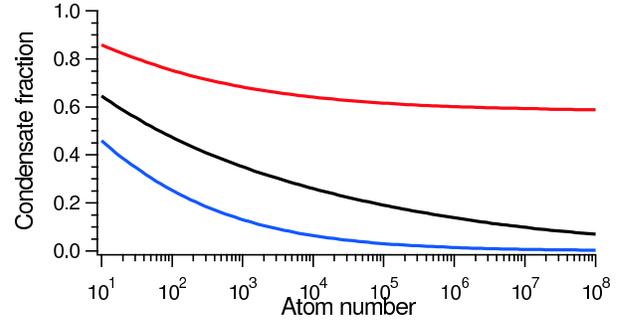}
\end{center}
\caption{Contribution of the ground-state on the peak density for,
from bottom to top, 1D, 2D and 3D images in function of the number
of trapped atoms. The clouds are at the transition temperature
$T^*_{ex}$ and the calculations use model $ex$. A 3D image would
give the density in all three dimensions of space \cite{nous}
whereas a 2D (resp. 1D) image corresponds to the density
integrated over one (resp. 2) dimension. For $N=10^4$ atoms the
ground-state contributes to $\sim 26\%$ in 2D images and $\sim
6\%$ in 1D images. In contrast to 3D image, the ground-state
contribution is very small for large atom number; it is not for
typical atom numbers accessible in experiments.}
 \label{fig5}
\end{figure}

We have shown results on the atomic density at the vicinity of the
transition temperature. Detection techniques consist rather on
1D-integrated density, corresponding to 2D absorption images, or
2D-integrated density \cite{BecHe}. One can show that, at
threshold, the 1D and 2D-integrated peak density of the
ground-state are vanishingly small for large atom numbers on
contrary to the non-integrated case. The peak 1D-integrated
density fraction behaves at threshold as $\sqrt{\tau}$ and the
2D-integrated peak density as $\tau$. For typical atom number this
is nevertheless not negligible. This is illustrated in
Fig.\ref{fig5} where is plotted the condensate peak density
fraction for 3D, 2D and 1D images of clouds at threshold. The
calculations use the model $ex$. At the transition temperature
$T^*_{ex}$, the ground-state contributes to more than $10\%$ for
$N<2500$ atoms in 1D images and for $N<8\;10^6$ atoms for 2D
images. It means that, even with the conventional technique of
absorption images, the effect should be experimentally observable
if interactions could be switched off using, for instance, the
magnetic tunability of the scattering length close to a Feshbach
resonance \cite{feshbach}.

Apart from the atomic density, two- and three-body inelastic loss
rates will also be affected and could be 20 to 30~\% higher than
predicted by model $sc_0$ around the transition temperature for
typical atom numbers. Finally, in most experimental set-ups, the
trapping potential is anisotropic and finite size effects are then
stronger. Indeed the term ${3\over 2}\tau$ in Eq.(\ref{equations
semi-classique}) should be replaced by ${3\over
2}\frac{\tilde\om}{\bar\om}\tau $, with
$\bar\om=(\prod\limits_i\om_i)^{1/3}$ the geometric mean and
$\tilde\om={1\over 3}\sum\limits_i\om_i$ the arithmetic mean
\cite{grossmann}. Whatever the anisotropy is, $\tilde\om$ is
always larger than $\om$, making the finite size contribution
stronger. To first order and if $k_BT^*_{ex}\gg\hbar\om_i$ for
$i=x,y$ and $z$, the ground-state contribution should be the same
since our high-N analysis does not depend on any anisotropy.

In conclusion, we have shown that the density of an ideal atomic
gas is dominated by the ground-state contribution near the
transition temperature. The inter-atomic interactions have been
neglected in our analysis and will modify our conclusions. With
repulsive interactions, the clouds tends to decrease its density
at the centre of the cloud whereas it tends to increase it with
attractive interactions. Previous calculations have treated
separately finite size and interactions effects, both corrections
being finally added \cite{stringari}. Since the ground-state has a
non-perturbative effect on the density, our analysis tends to
prove that both effects have to be investigated together. The
approach of Ref.\cite{krauth96} could in this respect provide
helpful informations. Feshbach resonances, which enable to tune
the interactions strength, constitute a powerful tool to check the
accuracy of the different theoretical models. Moreover, a full
three-dimensional density measurement would also be valuable; this
type of measurement is at the edge to be available in our
experiment on metastable helium in Orsay \cite{nous}.

\begin{acknowledgement}
We thank S. Giorgini for stimulating discussions. The Atom Optics
group of LCFIO is member of the Institut Francilien de Recherche
sur les Atomes Froids (IFRAF).
\end{acknowledgement}


\begin{thebibliography}{2}
\bibitem{resTc}
In Ref.\cite{pathria}, the transition temperature was indeed
$T_{sc}$ (see later in the text). The transition temperature
defined by the maximum of the second derivative of the condensate
fraction has been calculated for atom number in the range
$10^3-10^8$; the relative deviation is below $\sim 10^{-3}$ on the
transition temperature and $\sim 10^{-2}$ on the condensate peak
density fraction.
\bibitem{bose}
We use the usual definition of Bose functions
$g_a(x)=\sum\limits_{l=1}^\infty x^l/l^a$. We remind that
$g_a(1)=\zeta(a)$ with $\zeta()$ the Riemann Zeta function. Note
that $g_1(x)=-\ln(1-x)$ and ${d g_a\over d x}(x)=g_{a-1}(x)/x$.
\bibitem{calcul}
We use the result of Ref. \cite{robinson} for the calculation of
the Bose functions near the transition temperature. For the series
in model $ex$, the convergence is very slow but can easily be
accelerated. For instance, it is much better to write $N={z\over
1-z}+\sum\limits_{l=1}^\infty z^l({1\over 1-e^{-\tau l }}-1)$
because the second term converges for large $l$ because of the
$z^l$ part {\it and} because of the $({1\over 1-e^{-\tau l }}-1)$
part.
\end{thebibliography}
\end{document}